# A method of comparison between a force curve measured on a solvated surface and the solvation structure

**Ken-ichi Amano[a*], Kazuhiro Suzuki[b], Takeshi Fukuma[c], and Hiroshi Onishi[a]**

[a]*Department of Chemistry, Faculty of Science, Kobe University, Nada-ku, Kobe* 657-8501, *Japan*.

[b]*Department of Electronic Science and Engineering, Kyoto University, Katsura, Nishikyo, Kyoto 615-8510, Japan*.

[c]*Bio-AFM Frontier Research Center, Kanazawa University, Kakuma-machi, Kanazawa 920-1192, Japan*.

*Corresponding author: k-amano@gold.kobe-u.ac.jp

## 1. Introduction

Recent frequency modulated atomic force microscopy (FM-AFM) in liquid [1-5] can measure a force curve between a probe and a sample surface. The force curve is supposed to be the solvation structure formed on the sample surface in some cases, because its shape is generally oscillative and pitch of the oscillation is about the same as diameter of the solvent particle. However, it is not the solvation structure. It is just only an interaction force between the probe and sample surface. (Hereafter, we call the interaction force as a mean force.) In the present circumstances, unfortunately, a theoretical relation between the force curve and solvation structure is not clearly known. To date, only a simple relational expression between the force curve and solvation structure has been suggested by introducing a dot like probe (i.e., the probe is approximated to be a delta function which can penetrate a solvent particle until it arrives the *center point* of the solvent particle) [3]. To elucidate the relation between the force curve and solvation structure in more detail, we derive the relational expression by introducing a model probe with certain volume. The derivation is performed using the statistical mechanics of liquid [6], where influence of existence of the probe on the solvation structure formed on the sample surface is (implicitly) taken into account which is also an improved point against the previous study.

The relation between the force curve and solvation structure is studied in following theoretical condition: The probe and the sheet of the sample surface are immersed in the simple liquid. A usual simple liquid is composed of small hard-spheres or Lennard-Jones (LJ) spheres. The LJ spheres represent that each sphere interacts with LJ potential (e.g., octamethylcyclotetrasiloxane (OMCTS) and carbon tetrachloride ($CCl_4$) are typical models of LJ liquids). In some analyses [7-13], water is approximated to small hard-spheres or LJ liquid. When its number density is set to that of water, general property of the translational entropy of water can be captured. In addition, when its LJ attractive interaction is sufficiently strong, general property of the hydration energy can also be captured *in some degree*. Therefore, results of the present study with the simple liquid are not so limited results, which are thought to have some universality (i.e., the present results could be applied in the water AFM system as a first step study). The theoretical system is treated in the canonical ensemble and the system volume is supposed to be sufficiently large. A sheet of the sample surface is immersed in the solvent. The sheet and probe are treated as the solutes. Their orientations cannot be changed (are fixed), but three-dimensional placements can be changed. In the first half of the theoretical derivation, the tip apex (nano-cluster probe [14]) is treated, however, to connect the force curve and solvation structure within a simple relationship, an ideal probe is introduced. In the ideal probe, the solvent particle is attached to the tip apex of the probe and the other parts are neglected (i.e., the other parts do not have volumes and interactions with the solvent particles and the sheet). Consequently, the force curve and solvation structure are linked within a simple equation. In other words, the present study connects the force curve measured by the ideal probe and solvation structure. (Direct connection between the force curve measured by the nano-cluster probe and solvation structure is next study.)

The present study is always carried out in equilibrium state. Hence, the mean force between the sheet and probe is treated as that of conservative one, though the measured force is not exactly conservative force (however, it is nearly the conservative force in general). In the theoretical condition, the tip apex (nano-cluster probe) and ideal probe are assumed to be the probe models. That is, only a small part of the probe is considered. This assumption is considered to be valid, because it is demonstrated by the liquid AFM [4] that the only small part of the tip apex is important for the experimental result. In addition, since the probe of AFM is able to measure sample surface *at molecular resolution*, it is conceivable that the only small part of the tip apex with Angstrom seize has significant role in the measurement. Therefore, we have applied only the small part in the theoretical probe model.

In this paper, we explain the relation between the force curve and solvation structure in Chapter 2 and propose the method for comparing them in Chapter 3. The discussions about the force curve and solvation structure are done in a following background: The force curve is that obtained by the liquid AFM, and the solvation structure is that obtained by a calculation [3,14,15] or an experiment [16-19]. We carry out the present study by using statistical mechanics of liquid in equilibrium state.

## 2. Theory

Explanation of the method for comparison between the force curve and solvation structure is performed within the simple liquid in this paper. A usual simple liquid is composed of small hard-spheres or Lennard-Jones (LJ) spheres. The LJ spheres represent that each sphere interacts with LJ potential (e.g., octamethylcyclotetrasiloxane (OMCTS) and carbon tetrachloride ($CCl_4$) are typical models of LJ liquids). Let us consider a system of the canonical ensemble where the probe and the sheet of the sample surface are immersed in the solvent as the solutes. We do not consider whole volume of the probe. In the first half the probe is modeled as the nano-cluster probe [14], and in the last half it is modeled as the ideal probe. Although the sheet and probe are able to change their placements in the theory, their orientational variations are prohibited for the theoretical simplicity. External field acting on the system is set as zero. Conclusions obtained in the canonical ensemble are identical to that obtained in the grand canonical ensemble when the system volume is sufficiently large. Therefore, the theory is constructed in the canonical ensemble. (It is simpler for us to study the relation in the canonical ensemble rather than in the grand canonical one.) Here, we start derivation of the relational expression between the force curve and solvation structure by describing equations below. In the condition introduced above, the fundamental partition function ($Q_O$) can be written as

$$Q_O = \frac{\zeta^N}{N!}\int ...\int \exp\{-\beta U(\mathbf{r}_M{}', \mathbf{r}_P{}', \mathbf{r}_1, \mathbf{r}_2, ..., \mathbf{r}_N)\} d\mathbf{r}_M{}' d\mathbf{r}_P{}' d\mathbf{r}_1 d\mathbf{r}_2 ... d\mathbf{r}_N \ , \tag{1}$$

where $\zeta$ is expressed as

$$\zeta = (2\pi mkT)^{3/2} / h^3 \ . \tag{2}$$

Here, $\mathbf{r}$ represents three-dimensional vector that is expressed as $\mathbf{r}=(x,y,z)$ or $x\mathbf{i}+y\mathbf{j}+z\mathbf{k}$ where $\mathbf{i}$, $\mathbf{j}$, and $\mathbf{k}$ are unit vectors of $x$, $y$, and $z$-axes. $\int \mathrm{d}\mathbf{r}$ is volume integral, in another way, which is represented as $\iiint \mathrm{d}x\mathrm{d}y\mathrm{d}z$. Integrations of which are performed in the system of volume $V$. Characters of $N$, $\beta$, and $U$ represent the number of the solvent particles, 1 divided by "Boltzmann's constant ($k$) times temperature ($T$)", and internal energy, respectively. Subscripts M, P, and 1, 2, and $N$ represent the sheet of the sample surface, probe (with arbitrary shape), and each solvent particle, respectively. $\pi$, $m$, and $h$ are circle ratio, weight of the solvent particle, and Planck's constant, respectively. In this stage, the probe is supposed to be the nano-cluster probe. (Since the shape of the probe is not restricted in Eq. (1), the probe can have the shape of the ideal probe, too. In addition, the probe is supposed to be distinguishable from the solvent particles even when the probe is same as the solvent particle of the ideal probe.) In Eq. (1), integrated terms of kinetic momenta for the sheet and probe are neglected, because their positions are changed artificially. That is, they do not change their positions with their kinetic momenta. Next, we shall describe the partition function where the sheet and probe are fixed at $\mathbf{r}_M$ and $\mathbf{r}_P$, respectively, which ($Q$) is expressed as

$$Q(\mathbf{r}_M, \mathbf{r}_P) = \frac{\zeta^N}{N!}\int ...\int \exp\{-\beta U(\mathbf{r}_M, \mathbf{r}_P, \mathbf{r}_1, \mathbf{r}_2, ..., \mathbf{r}_N)\} d\mathbf{r}_1 d\mathbf{r}_2 ... d\mathbf{r}_N \ . \tag{3}$$

Eq. (3) is very important partition function, because free energy ($F$) of the system where the sheet and probe are fixed at certain positions can readily be obtained by using a basic equation: $F=-kT\ln Q$.

To obtain the relation between the force curve and solvation structure, we shall see insides of the mean force between the sheet and probe ($\mathbf{f}_{\mathrm{MP}}$). It is written as

$$\mathbf{f}_{\mathrm{MP}}(\mathbf{r}_{\mathrm{M}},\mathbf{r}_{\mathrm{P}}) \equiv -\frac{\partial \Phi_{\mathrm{MP}}(\mathbf{r}_{\mathrm{M}},\mathbf{r}_{\mathrm{P}})}{\partial \mathbf{r}_{\mathrm{P}}} = -\frac{\partial\{F(\mathbf{r}_{\mathrm{M}},\mathbf{r}_{\mathrm{P}}) - F(\infty,\infty)\}}{\partial \mathbf{r}_{\mathrm{P}}}, \tag{4}$$

where $\Phi_{\mathrm{MP}}$ is potential of mean force between the sheet and the probe and $(\infty,\infty)$ means the sample surface and probe are infinitely separated [12,13]. Partial differentiation of vector $\mathbf{r}$ denotes $\partial/\partial\mathbf{r}=(\partial/\partial x)\mathbf{i}+(\partial/\partial y)\mathbf{j}+(\partial/\partial z)\mathbf{k}$. It is defined in Eq. (4) that when a value of $\mathbf{f}_{\mathrm{MP}}$ of $i$-component ($i=x$, $y$, or $z$) is positive the probe feels force whose direction is same as $i$-axis, while when the value is negative the probe feels force whose direction is opposite to $i$-axis. Using the basic equation, Eq. (4) is rewritten as

$$\mathbf{f}_{\mathrm{MP}}(\mathbf{r}_{\mathrm{M}},\mathbf{r}_{\mathrm{P}}) = -\frac{\partial}{\partial \mathbf{r}_{\mathrm{P}}}(-kT\ln Q) = kT\frac{1}{Q}\frac{\partial Q}{\partial \mathbf{r}_{\mathrm{P}}}. \tag{5}$$

Next, we shall see expressions for the pair distribution function between the sheet and probe ($g_{\mathrm{MP}}$), which is expressed as

$$g_{\mathrm{MP}}(\mathbf{r}_{\mathrm{M}},\mathbf{r}_{\mathrm{P}}) \equiv \frac{\rho_{\mathrm{MP}}(\mathbf{r}_{\mathrm{M}},\mathbf{r}_{\mathrm{P}})}{\rho_{\mathrm{M}}\cdot\rho_{\mathrm{P}}} = \frac{1}{\rho_{\mathrm{M}}\cdot\rho_{\mathrm{P}}}\left\langle \delta(\mathbf{r}_{\mathrm{M}}-\mathbf{r}_{\mathrm{M}}{'})\delta(\mathbf{r}_{\mathrm{P}}-\mathbf{r}_{\mathrm{P}}{'})\right\rangle, \tag{6}$$

where $\rho_{\mathrm{MP}}$, $\rho_{\mathrm{M}}$, and $\rho_{\mathrm{P}}$ represent pair density distribution between the sheet and probe, bulk densities of the sheet and probe, respectively ($\rho_{\mathrm{M}}$ and $\rho_{\mathrm{P}}$ are constants). $<X>$ represents an ensemble average of $X$. Hence, Eq. (6) is calculated to be

$$g_{\mathrm{MP}}(\mathbf{r}_{\mathrm{M}},\mathbf{r}_{\mathrm{P}}) = \frac{Q(\mathbf{r}_{\mathrm{M}},\mathbf{r}_{\mathrm{P}})}{\rho_{\mathrm{M}}\rho_{\mathrm{P}}Q_{\mathrm{O}}} . \tag{7}$$

Thus, $kT(\partial/\partial\mathbf{r}_{\mathrm{P}})\ln(g_{\mathrm{MP}})$ can be written as [20]

$$kT\frac{\partial}{\partial\mathbf{r}_{\mathrm{P}}}\ln g_{\mathrm{MP}}(\mathbf{r}_{\mathrm{M}},\mathbf{r}_{\mathrm{P}}) = kT\frac{1}{Q}\frac{\partial Q}{\partial\mathbf{r}_{\mathrm{P}}} . \tag{8}$$

Since right-hand sides of Eqs. (5) and (8) are the same, $\mathbf{f}_{\mathrm{MP}}$ and $g_{\mathrm{MP}}$ have a following relationship,

$$\mathbf{f}_{\mathrm{MP}}(\mathbf{r}_{\mathrm{M}},\mathbf{r}_{\mathrm{P}}) = kT\frac{\partial}{\partial\mathbf{r}_{\mathrm{P}}}\ln g_{\mathrm{MP}}(\mathbf{r}_{\mathrm{M}},\mathbf{r}_{\mathrm{P}}) . \tag{9}$$

The final aim of this letter is not finding of the relational expression between $\mathbf{f}_{\mathrm{MP}}$ and $g_{\mathrm{MP}}$, but that between $\mathbf{f}_{\mathrm{MP}}$ and $g_{\mathrm{MS}}$ ($g_{\mathrm{MS}}$ is the pair distribution function between the sheet and solvent in which a subscript S denotes the solvent). Hence, we shall see the relation between $\mathbf{f}_{\mathrm{MP}}$ and $g_{\mathrm{MS}}$ by introducing the ideal probe. In the case of the ideal probe, *potential* of the mean force between the sheet and ideal probe is the same as that between the sheet and the solvent particle, which is written in an equation form that

$$\Phi_{\mathrm{MP*}} = \Phi_{\mathrm{MS}} . \tag{10}$$

Here, the subscript P* represents the ideal probe. The equality written in Eq. (10) is surely obvious one. For the reliability of the Eq. (10), however, we have confirmed the fact by using Ornstein–Zernike theory coupled by HNC closure, and we have concluded that it is soundly true. By the way, according to Eqs. (4) and (9), potential of the mean force ($\Phi_{\mathrm{MP}}$) is calculated to be

$$\Phi_{\mathrm{MP}}(\mathbf{r}_{\mathrm{M}},\mathbf{r}_{\mathrm{P}})=\int_{C}\mathbf{f}_{\mathrm{MP}}(\mathbf{r}_{\mathrm{M}},\mathbf{r}_{\mathrm{P}}{}')\cdot d\mathbf{r}_{\mathrm{P}}{}'=-kT\ln g_{\mathrm{MP}}(\mathbf{r}_{\mathrm{M}},\mathbf{r}_{\mathrm{P}}), \qquad (11)$$

where $C$ represents a curvilinear integral and $\mathrm{d}\mathbf{r}_{\mathrm{P}}'=\mathrm{d}x_{\mathrm{P}}'\mathbf{i}+\mathrm{d}y_{\mathrm{P}}'\mathbf{j}+\mathrm{d}z_{\mathrm{P}}'\mathbf{k}$. The integral range is from $\mathbf{r}_{\mathrm{P}}(=x_{\mathrm{P}}\mathbf{i}+y_{\mathrm{P}}\mathbf{j}+z_{\mathrm{P}}\mathbf{k})$ to $\infty\mathbf{i}+\infty\mathbf{j}+\infty\mathbf{k}$ or from $\mathbf{r}_{\mathrm{P}}$ to $x_{\mathrm{P}}\mathbf{i}+y_{\mathrm{P}}\mathbf{j}+\infty\mathbf{k}$, etc. Since equilibrium state is considered here, the calculation result of the curvilinear integral does not depend on its integral path (i.e., the calculation result depends on the start and end of the integral). The pair distribution function between the sheet and probe ($g_{\mathrm{MP}}$) is simply expressed as

$$g_{\mathrm{MP}}(\mathbf{r}_{\mathrm{M}},\mathbf{r}_{\mathrm{P}})=\exp\{-\beta\Phi_{\mathrm{MP}}(\mathbf{r}_{\mathrm{M}},\mathbf{r}_{\mathrm{P}})\}. \qquad (12)$$

Referring and combining Eqs. (10) and (12), following relation can be written in the case of the ideal probe:

$$g_{\mathrm{MP}*}=\exp(-\beta\Phi_{\mathrm{MP}*})=\exp(-\beta\Phi_{\mathrm{MS}})=g_{\mathrm{MS}}. \qquad (13)$$

Eq. (13) represents that when the probe is ideal one, the pair distribution function between the sheet and ideal probe corresponds to that between the sheet and solvent particle. The above process is a simpler way to find the relational expression between $g_{\mathrm{MP}*}$ and $g_{\mathrm{MS}}$. This is because if $g_{\mathrm{MP}*}$ and $g_{\mathrm{MS}}$ are compared straightforwardly (without the supposed equality of Eq. (10)), relation between following two equations must be explored.

$$g_{\mathrm{MP}*}(\mathbf{r}_{\mathrm{M}},\mathbf{r}_{\mathrm{P}*})=\frac{1}{\rho_{\mathrm{M}}\rho_{\mathrm{P}*}}\cdot\frac{1}{Q_{\mathrm{O}}}\frac{\zeta^{N}}{N!}\int\cdots\int\exp\{-\beta U(\mathbf{r}_{\mathrm{M}},\mathbf{r}_{\mathrm{P}*},\mathbf{r}_{1},\mathbf{r}_{2},...,\mathbf{r}_{N})\}d\mathbf{r}_{1}d\mathbf{r}_{2}...d\mathbf{r}_{N}, \qquad (14)$$

$$g_{\mathrm{MS}}(\mathbf{r}_{\mathrm{M}},\mathbf{r}_{\mathrm{S}})=\frac{1}{\rho_{\mathrm{M}}\rho_{\mathrm{S}}}\cdot\frac{1}{Q_{\mathrm{A}}}\frac{\zeta^{N}}{(N-1)!}\int\cdots\int\exp\{-\beta U_{\mathrm{A}}(\mathbf{r}_{\mathrm{M}},\mathbf{r}_{\mathrm{S}},\mathbf{r}_{2},...,\mathbf{r}_{N})\}d\mathbf{r}_{2}...d\mathbf{r}_{N}, \qquad (15)$$

where $U_\mathrm{A}$ represents internal energy in the absence of the probe, and $Q_\mathrm{A}$ is expressed as

$$Q_\mathrm{A} = \frac{\zeta^N}{N!}\int \ldots \int \exp(-\beta U_\mathrm{A}) d\mathbf{r}_\mathrm{M} d\mathbf{r}_1 d\mathbf{r}_2 \ldots d\mathbf{r}_N \ . \tag{16}$$

There are several differences between Eqs. (14) and (15), however, fortunately, Eqs. (14) and (15) are proven to be (fairly) the same (see Appendix A).

Finally, the relational expression between $\mathbf{f}_\mathrm{MP}$ and $g_\mathrm{MS}$ is obtained by connecting Eqs. (9) and (13) as follows:

$$\mathbf{f}_\mathrm{MP}(\mathbf{r}_\mathrm{M}, \mathbf{r}_\mathrm{P})\big|_{\mathrm{P}\to\mathrm{P}*} = \mathbf{f}_{\mathrm{MP}*}(\mathbf{r}_\mathrm{M}, \mathbf{r}_{\mathrm{P}*})\big|_{\mathbf{r}_{\mathrm{P}*}\to\mathbf{r}_\mathrm{S}} = \mathbf{f}_{\mathrm{MP}*}(\mathbf{r}_\mathrm{M}, \mathbf{r}_\mathrm{S}) = kT \frac{\partial}{\partial \mathbf{r}_\mathrm{S}} \ln g_\mathrm{MS}(\mathbf{r}_\mathrm{M}, \mathbf{r}_\mathrm{S}) \ . \tag{17}$$

Here, P→P* represents the probe with arbitrary shape (e.g., the nano-cluster probe) is changed to the ideal probe. The change can readily be done in the theoretical system, because the shape of the probe is not specified in the function of $U$. The replacement $\mathbf{r}_{\mathrm{P}*} \to \mathbf{r}_\mathrm{S}$ means that only the character is replaced from $\mathbf{r}_{\mathrm{P}*}$ to $\mathbf{r}_\mathrm{S}$ (i.e., its vector value is not changed). If the measured mean force is conservative force and the probe is the ideal probe, the force curve and solvation structure are connected by the simple equation. Although existence of the ideal probe in the vicinity of the sheet also deforms the solvation structure on the sheet, the force curve and solvation structure are simply connected. This is one of the benefits of the present study.

## 3. Discussion

In the present chapter, a method for comparing between the force curve and solvation structure is discussed. In the real AFM system, in general, the probe is always on the upper side of the sample surface and measured mean forces are that (almost) along $z$-axis. That's why, we include the above two general aspects in the discussion. That is, following two settings are included: $z_M<z_P$ and $\mathbf{f}_{MP}\cdot \mathbf{k}=f_{MPz}$ ($f_{MPz}$ is the mean force between the sheet and probe along $z$-axis). The procedure for the comparison between $f_{MPz}$ and $g_{MS}$ is as follows:

(I) Measure $f_{MPz}$ by using AFM in the (simple) liquid.

(II) Obtain $g_{MS}$ from a calculation or an experiment.

(III) Calculate $f_{MP^*z}$ by substituting the $g_{MS}$ into Eq. (18), where it is hypothecated that the ideal probe is used in the system of (II) (not in the AFM system of (I)).

(IV) Compare shapes of the $f_{MPz}$ and $f_{MP^*z}$. When $f_{MPz}$ is well accorded with $f_{MP^*z}$, the probe used in the AFM system is considered to be an almost the ideal probe. In this case, solvation structure can approximately be estimated from the measured $f_{MPz}$ using Eq. (19). On the other hand, when the $f_{MPz}$ is not similar to the $f_{MP^*z}$, it exposes that the probe used in the AFM system is clearly different from the ideal probe.

$$kT\frac{\partial}{\partial z_S}\ln g_{MS}(\mathbf{r}_M,\mathbf{r}_S)\bigg|_{\mathbf{r}_S\to\mathbf{r}_{P^*}} = f_{MP^*z}(\mathbf{r}_M,\mathbf{r}_{P^*})\,. \qquad (18)$$

$$\exp\left\{-\beta\int_{z_P}^{\infty} f_{MPz}(\mathbf{r}_M,x_P,y_P,z_P{}')dz_P{}'\right\}\bigg|_{\mathbf{r}_P\to\mathbf{r}_S} \approx g_{MS}(\mathbf{r}_M,\mathbf{r}_S)\,. \qquad (19)$$

The replacements of $\mathbf{r}_S\to\mathbf{r}_{P^*}$ and $\mathbf{r}_P\to\mathbf{r}_S$ in Eqs. (18) and (19) mean that only the characters are replaced (i.e., the vector values are not changed).

If it had been known in the first place that the probe used in the real AFM system was almost the ideal probe, the solvation structure can be estimated from the force curve through Eq. (19). It implies that development of a nano-technology which can fabricates the probe with almost ideal one is a key technology for obtaining the solvation structure from the liquid AFM.

## 4. Conclusions

In summary, we have shown the relational expression between $\mathbf{f}_{\mathrm{MP}}$ ($f_{\mathrm{MPz}}$) and $g_{\mathrm{MS}}$ by introducing the ideal probe. The method for comparing the force curve and solvation structure has been proposed. The relation between $\mathbf{f}_{\mathrm{MP}}$ and $g_{\mathrm{MS}}$, which is represented as $\mathbf{f}_{\mathrm{MP}} \leftrightarrow g_{\mathrm{MS}}$, has been derived by following a route: $\mathbf{f}_{\mathrm{MP}} \leftrightarrow g_{\mathrm{MP}} \leftrightarrow g_{\mathrm{MP*}} \leftrightarrow \Phi_{\mathrm{MP*}} \leftrightarrow \Phi_{\mathrm{MS}} \leftrightarrow g_{\mathrm{MS}}$. The relation can also be derived by following a straightforward route: $\mathbf{f}_{\mathrm{MP}} \leftrightarrow g_{\mathrm{MP}} \leftrightarrow g_{\mathrm{MP*}} \leftrightarrow g_{\mathrm{MS}}$ (see Appendix A). The latter route is a strict route compared with the former route, because the former route has introduced a hypothesis that $\Phi_{\mathrm{MP*}}$ is equal to $\Phi_{\mathrm{MS}}$. If the force curve along $z$-axis is measured using the ideal probe, the force curve and solvation structure are connected in the form of Eq. (18). In other words, the force curve and solvation structure are connected by the simple equation if the probe is the ideal one, although existence of the ideal probe in the vicinity of the sheet also deforms the solvation structure on the sheet. This is the most important conclusion of the present study.

The introduction of the ideal probe has been readily performed in the derivation. This is because, in the theory of the first half, the shape of the probe is not restricted and it can take arbitrary shape. In this case, the change from the probe with arbitrary shape to the ideal probe can readily be done. Hence, $g_{\mathrm{MP}}$ and $g_{\mathrm{MP*}}$ are immediately connected in the derivation.

The relational expression between $\mathbf{f}_{\mathrm{MP}}$ and $g_{\mathrm{MS}}$ (Eq. (17)) indicates that the ideal probe receives a vector force that moves the ideal probe to the local maximum of the $g_{MS}$ (i.e., it indicates that the ideal probe is stabilized at the local maximum of the $g_{MS}$). This is reasonable, because the ideal probe is fairly the solvent particle. However, there remains a question: If the probe is a realistic probe (e.g., the nano-cluster probe), how does the force act on the probe? This question has not been solved in the present paper, however, if the probe is nearly the identical probe such a concern is minor concern. It is supposed that behavior of the nearly identical probe is similar to that of the ideal one. Actually, we have briefly checked the supposition by making use of an integral equation theory (Ornstein-Zernike equation coupled by hypernetted-chain closure) [6-13], though the result is not shown here. The discussion about the behavior of a probe with completely different property and shape against the ideal probe cannot be done in the present paper, which is one of our next challenges.

In the real AFM experiment, most of the probes are not identical one. This fact requires another method in comparison between the force curve and solvation structure. The alternative method is transformation of the measured force curve into the solvation structure, and the transformed solvation structure is compared with the solvation structure obtained by a calculation or an experiment. Recently, K. Amano [21] has proposed the method for calculating solvation structure from the measured force curve within one-dimensional model system. In the method, a sufficiently large sphere is modeled as the sample surface and a sphere with certain diameter is modeled as the probe. The transformation can be done even when the probe is either (highly) solvophilic or solvophobic, which is a different point against Eq. (19). However, there are problems in the transformation method. The method is restricted in the one-dimensional model system and shapes of the models of the sample surface and probe are fixed in spherical shapes. Solving of the problems and development of the transformation method into the three-dimensional model system are also next challenges of us.

As suggested in Chapter 3, fabrication of the probe with almost ideal one is a key nano-technology for the liquid AFM, because it is considered that the nearly identical probe can get reasonable information about the solvation structure according to Eq. (19). Then, we remark that it should be studied beforehand by a simulation that what kind of the probe is the most identical probe within the commercially available probes. We believe *such kinds of studies* provide significant information for fabrication of the nearly identical probe.

In the near future, it is likely that the time for a simulation of the mean force between the sample surface and *the probe with arbitrary shape* is shortened much. It enables us to compare the measured and simulated force curves easier. However, this comparison does not provide the information about the solvation structure purely formed on the sample surface whose structure is not sandwiched between the surfaces of the sample and probe. To extract the information about the solvation structure from the measurement, it is imperative to *theoretically* capture the relation between them. Therefore, we have derived the relational expression between them. We believe that this work deepens understanding of the mean force measured by the AFM in liquid and sheds light on the measurement of the solvation structure on the sample surface.

## Acknowledgements

We appreciate Masahiro Kinoshita (Kyoto University) and Hiraku Oshima (Kyoto University) for helpful advices. We thank Kei Kobayashi (Kyoto University) and Shinichiro Ido (Kyoto University) for useful comments. This work was supported by Grant-in-Aid for JSPS (Japan Society for the Promotion of Science) fellows and Foundation of Advanced Technology Institute.

## Appendix A. Comparison between $g_{\mathrm{MP*}}$ and $g_{\mathrm{MS}}$

The $g_{\mathrm{MP*}}$ and $g_{\mathrm{MS}}$ are straightforwardly compared here. Describing the contents of the fundamental partition function ($Q_{\mathrm{O}}$), $g_{\mathrm{MP*}}$ is rewritten as

$$g_{\mathrm{MP*}}(\mathbf{r}_{\mathrm{M}},\mathbf{r}_{\mathrm{P*}}) = \frac{1}{\rho_{\mathrm{M}}\rho_{\mathrm{P*}}} \cdot \frac{\dfrac{\zeta^N}{N!}\int ... \int \exp\{-\beta U(\mathbf{r}_{\mathrm{M}},\mathbf{r}_{\mathrm{P*}},\mathbf{r}_1,\mathbf{r}_2,...,\mathbf{r}_N)\}d\mathbf{r}_1 d\mathbf{r}_2 ... d\mathbf{r}_N}{\dfrac{\zeta^N}{N!}\int ... \int \exp\{-\beta U(\mathbf{r}_{\mathrm{M}}',\mathbf{r}_{\mathrm{P*}}',\mathbf{r}_1,\mathbf{r}_2,...,\mathbf{r}_N)\}d\mathbf{r}_{\mathrm{M}}' d\mathbf{r}_{\mathrm{P*}}' d\mathbf{r}_1 d\mathbf{r}_2 ... d\mathbf{r}_N}, \tag{20}$$

and which is calculated as

$$g_{\mathrm{MP*}}(\mathbf{r}_{\mathrm{M}},\mathbf{r}_{\mathrm{P*}}) = V^2 \cdot \frac{\int ... \int \exp\{-\beta U(\mathbf{r}_{\mathrm{M}},\mathbf{r}_{\mathrm{P*}},\mathbf{r}_1,\mathbf{r}_2,...,\mathbf{r}_N)\}d\mathbf{r}_1 d\mathbf{r}_2 ... d\mathbf{r}_N}{\int ... \int \exp\{-\beta U(\mathbf{r}_{\mathrm{M}}',\mathbf{r}_{\mathrm{P*}}',\mathbf{r}_1,\mathbf{r}_2,...,\mathbf{r}_N)\}d\mathbf{r}_{\mathrm{M}}' d\mathbf{r}_{\mathrm{P*}}' d\mathbf{r}_1 d\mathbf{r}_2 ... d\mathbf{r}_N}, \tag{21}$$

where $V$ represents volume of the system. If the ideal probe is located at $\mathbf{r}_{\mathrm{S}}$ ($\mathbf{r}_{\mathrm{P*}}\rightarrow\mathbf{r}_{\mathrm{S}}$) and characters of $\mathbf{r}_{\mathrm{P*}}$' at denominator are alternated to $\mathbf{r}_0$ ($\mathbf{r}_{\mathrm{P*}}'\rightarrow\mathbf{r}_0$), Eq. (21) is rewritten as

$$g_{\mathrm{MP*}}(\mathbf{r}_{\mathrm{M}},\mathbf{r}_{\mathrm{S}}) = V^2 \cdot \frac{\int ... \int \exp\{-\beta U(\mathbf{r}_{\mathrm{M}},\mathbf{r}_{\mathrm{S}},\mathbf{r}_1,\mathbf{r}_2,...,\mathbf{r}_N)\}d\mathbf{r}_1 d\mathbf{r}_2 ... d\mathbf{r}_N}{\int ... \int \exp\{-\beta U(\mathbf{r}_{\mathrm{M}}',\mathbf{r}_0,\mathbf{r}_1,\mathbf{r}_2,...,\mathbf{r}_N)\}d\mathbf{r}_{\mathrm{M}}' d\mathbf{r}_0 d\mathbf{r}_1 d\mathbf{r}_2 ... d\mathbf{r}_N}. \tag{22}$$

Since the probe considered here is the ideal one, Eq. (22) is converted to be

$$g_{\mathrm{MP*}}(\mathbf{r}_{\mathrm{M}},\mathbf{r}_{\mathrm{S}}) = V^2 \cdot \frac{\int ... \int \exp(-\beta U_{\mathrm{A}}(\mathbf{r}_{\mathrm{M}},\mathbf{r}_{\mathrm{S}},\mathbf{r}_2,...,\mathbf{r}_{N+1}))d\mathbf{r}_2 ... d\mathbf{r}_{N+1}}{\int ... \int \exp\{-\beta U_{\mathrm{A}}(\mathbf{r}_{\mathrm{M}}',\mathbf{r}_1,\mathbf{r}_2,...,\mathbf{r}_{N+1})\}d\mathbf{r}_{\mathrm{M}}' d\mathbf{r}_1 d\mathbf{r}_2 ... d\mathbf{r}_{N+1}}. \tag{23}$$

Next, we shall see the contents of $g_{\mathrm{MS}}$ by referring Eqs. (15) and (16),

$$g_{\mathrm{MS}}(\mathbf{r}_{\mathrm{M}},\mathbf{r}_{\mathrm{S}}) = \frac{1}{\rho_{\mathrm{M}}\rho_{\mathrm{S}}} \cdot \frac{\frac{\zeta^{N}}{(N-1)!}\int\ldots\int \exp(-\beta U_{\mathrm{A}}(\mathbf{r}_{\mathrm{M}},\mathbf{r}_{\mathrm{S}},\mathbf{r}_{2},\ldots,\mathbf{r}_{N}))d\mathbf{r}_{2}\ldots d\mathbf{r}_{N}}{\frac{\zeta^{N}}{N!}\int\ldots\int \exp\{-\beta U_{\mathrm{A}}(\mathbf{r}_{\mathrm{M}}',\mathbf{r}_{1},\mathbf{r}_{2},\ldots,\mathbf{r}_{N})\}d\mathbf{r}_{\mathrm{M}}'d\mathbf{r}_{1}d\mathbf{r}_{2}\ldots d\mathbf{r}_{N}}. \tag{24}$$

Since respective $\rho_{\mathrm{M}}$ and $\rho_{\mathrm{S}}$ are $(1/V)$ and $(N/V)$, Eq. (24) is calculated to be

$$g_{\mathrm{MS}}(\mathbf{r}_{\mathrm{M}},\mathbf{r}_{\mathrm{S}}) = V^{2} \cdot \frac{\int\ldots\int \exp(-\beta U_{\mathrm{A}}(\mathbf{r}_{\mathrm{M}},\mathbf{r}_{\mathrm{S}},\mathbf{r}_{2},\ldots,\mathbf{r}_{N}))d\mathbf{r}_{2}\ldots d\mathbf{r}_{N}}{\int\ldots\int \exp\{-\beta U_{\mathrm{A}}(\mathbf{r}_{\mathrm{M}}',\mathbf{r}_{1},\mathbf{r}_{2},\ldots,\mathbf{r}_{N})\}d\mathbf{r}_{\mathrm{M}}'d\mathbf{r}_{1}d\mathbf{r}_{2}\ldots d\mathbf{r}_{N}}. \tag{25}$$

As a result, it is revealed that $g_{\mathrm{MP}*}$ is (fairly) equal to $g_{\mathrm{MS}}$ when $N$ is sufficiently large ($1 \ll N$) by comparing Eqs. (23) and (25). Range of $N$ is discussed as follows: If an infinite number of the solvent particles exist in the system with volume $V$, the internal energy becomes infinite due to the extremely high crowding of the solvent particles. It implies that the system cannot exist completely. That is, when $N$ is infinite, the system loses the physical meaning (i.e., the ensemble is neither fluid nor solid). Therefore, the range of $N$ is considered to be $1 \ll N < \infty$.

1st Submission: Sun., 9 Dec. 2012 08:01:01 EST.

2nd Submission: Wed., 19 Dec. 2012 EST.

(The main text and references were modified.)

3rd Submission: Sun., 23 Dec. 2012 EST.

(The appendix was inserted.)

4th Submission: Sat., 29 Dec. 2012 EST.

(The main text was modified.)

Update of 4th Submission: Mon., 31 Dec. 2012 EST.

(The main text was modified.)

5th Submission: Fri., 1 Feb. 2013 EST.

(Coauthors and additional acknowledgements are added.)

6th Submission: Wed, 6 Mar. 2013 EST.

(The main text and acknowledgements are modified.)